\newif\ifdouble
\newif\ifsingle
\newif\ifchange
\doubletrue

\documentclass[sigconf]{acmart}

\usepackage{dblfloatfix} 
\usepackage{booktabs} 
\usepackage{arydshln}
\usepackage{subfig}
\AtBeginDocument{%
  \providecommand\BibTeX{{%
    \normalfont B\kern-0.5em{\scshape i\kern-0.25em b}\kern-0.8em\TeX}}}

\copyrightyear{2022}
\acmYear{2022}
\setcopyright{rightsretained}
\acmConference[UIST '22 Adjunct]{The Adjunct Publication of the 35th Annual ACM Symposium on User Interface Software and Technology}{October 29-November 2, 2022}{Bend, OR, USA}
\acmBooktitle{The Adjunct Publication of the 35th Annual ACM Symposium on User Interface Software and Technology (UIST '22 Adjunct), October 29-November 2, 2022, Bend, OR, USA}
\acmDOI{10.1145/3526114.3561350}
\acmISBN{978-1-4503-9321-8/22/10}

\newcommand{\system}{UltraBots}

\begin{document}
\pagenumbering{arabic}
\pagestyle{plain}
\title{\system{}: Large-Area Mid-Air Haptics for VR with Robotically Actuated Ultrasound Transducers}

\author{Mehrad Faridan}
\affiliation{%
  \institution{University of Calgary}
  \streetaddress{Department of Computer Science}
  \city{Calgary}
  \country{Canada}}
\email{mehrad.faridan1@ucalgary.ca}

\author{Marcus Friedel}
\affiliation{%
  \institution{University of Calgary}
  \streetaddress{Department of Computer Science}
  \city{Calgary}
  \country{Canada}}
\email{marcus.friedel@ucalgary.ca}

\author{Ryo Suzuki}
\affiliation{%
  \institution{University of Calgary}
  \streetaddress{Department of Computer Science}
  \city{Calgary}
  \country{Canada}}
\email{ryo.suzuki@ucalgary.ca}

\renewcommand{\shortauthors}{Anonymous, et al.}

\begin{abstract}
We introduce \system{}, a system that combines ultrasound haptic feedback and robotic actuation for large-area mid-air haptics for VR. Ultrasound haptics can provide precise mid-air haptic feedback and versatile shape rendering, but the interaction area is often limited by the small size of the ultrasound devices, restricting the possible interactions for VR. To address this problem, this paper introduces a novel approach that combines robotic actuation with ultrasound haptics. More specifically, we will attach ultrasound transducer arrays to tabletop mobile robots or robotic arms for scalable, extendable, and translatable interaction areas. We plan to use Sony Toio robots for 2D translation and/or commercially available robotic arms for 3D translation. Using robotic actuation and hand tracking measured by a VR HMD (ex: Oculus Quest), our system can keep the ultrasound transducers underneath the user's hands to provide on-demand haptics. We demonstrate applications with workspace environments, medical training, education and entertainment. 
\end{abstract}



\maketitle

\section{Introduction}

Haptics in VR can enhance user experience by moving beyond the sense of vision to enable users to interact using physical channels and to internalize and understand information via the sense of touch. HCI research has investigated many types of haptic devices, including controller-based and wearable haptics. However, these haptic devices often require special pre-configuration or limited generalizability of shape-rendering. In contrast, ultrasound haptics can address both accessibility and generalizability problems. For example, Ultraleap's mid-air haptics device \cite{ultraleap} uses ultrasound transducers for haptic tactile feedback, shape rendering and more. Mouth Haptics \cite{shen2022mouthhaptics} uses ultrasound haptics for mouth based force-feedback. As we can see, ultrasound-based haptics hold much potential for haptic approaches for the future of VR development. 

\begin{figure}[t!]
\centering
\includegraphics[width=1\linewidth]{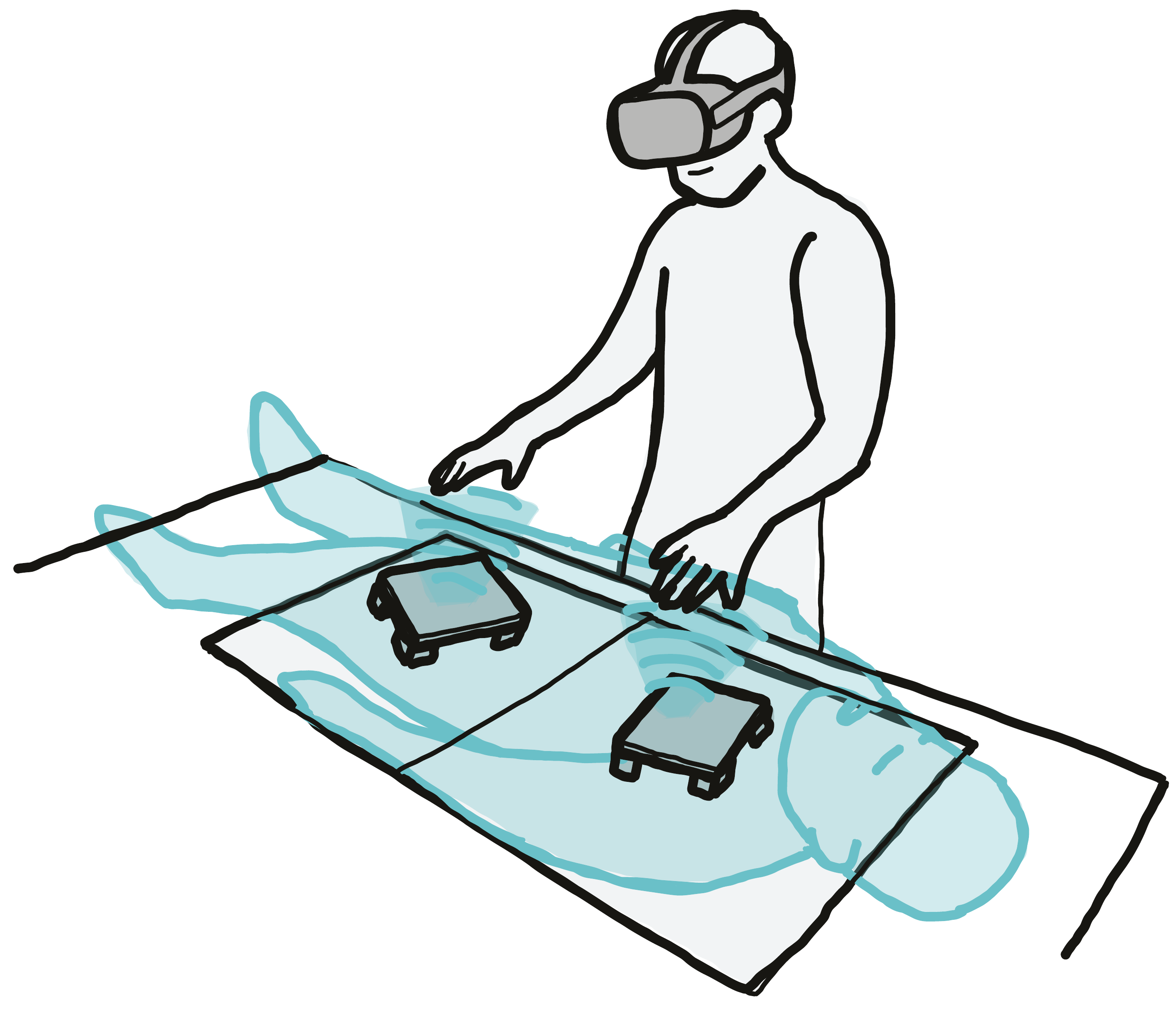}
\caption{The \system{} overview: robotic positioning of ultrasound transducers for large-area mid-air haptics.}
\label{fig:teaser}
\end{figure}

However, one of the limitations of the current ultrasound haptics is a limited and fixed interaction area. For example, Ultraleap haptics can only cover a 63cm $\times$ 48cm $\times$ 48cm area, which is inherently limiting due to the size of the ultrasound transducers. Because of that, the area in which haptic feedback can be rendered is both small and stationary which limits many types of applications and interactions for VR. On the other hand, due to the expensive cost of ultrasound transducers, scaling their size  with additional devices is impractical. For example, the cost of a single Ultraleap Stratos Inspire/Explore \cite{ultraleap} is over 5,000 USD which implies that covering a 2 meter by 2 meter area with ultrasound transducers would cost at least 20,000 USD. In addition, the devices have a limited rendering height, limiting their ability to render haptic feedback in a 3D space.

To address this problem, this paper introduces \system{}, a new approach for large-area ultrasound haptics leveraging robotic actuation. The core idea of our approach is to combine ultrasound haptics with robots such as tabletop mobile robots or robotic arms to cover a large, scalable and flexible interaction area, as shown in Figure 1. Ultrasound haptics can provide precise, high-resolution haptic rendering for specific areas, whereas the attached robots can move based on the users hands and body position, to cover larger areas when required. To demonstrate our concept, we leverage Sony Toio robots, a tabletop wheeled mobile robot that moves along 2D surfaces. We attach ultrasound transducers on top of the Toio robots move the transducers over the area of a 2D surface. Based on the hand and position tracking, measured by a VR HMD (ex: Oculus Quest), we move these robots underneath the users hands to provide haptic feedback on-demand. This approach can also be scalable, allowing multiple robots to render haptic sensations for both hands simultaneously.


Additionally, This idea can also be generalized for other robotic platforms, like robotic arms. We are also interested in implementing this principle using robotic arms to cover 3D space. We demonstrate this concept through various applications, including medical training, workspace environment, education and entertainment.

Finally, this paper contributes to the following:
\begin{enumerate}
\item An approach for large-area mid-air haptics by combining ultrasound haptic transducors and robotic actuation. 
\item \system{}, a system that leverages Sony Toio robots for 2D translation of ultrasound transducors for accurate and precise mid-air haptics and an Oculus Quest for user position and hand tracking.
\item Application scenarios which demonstrate the possibility of large-area ultrasound haptic interactions.
\end{enumerate}

\section{Related Work}

Ultrasound is one of the three primary approaches for mid-air haptics, along with air-jets and laser-based systems \cite{Bermejo2021HapticTechnologySurvey}. Ultrasound haptics have been researched in many facets within HCI. \cite{Marti2022MidAirShapeRecognition, Long2014RenderingVolumetricHapticShapes} use ultrasound haptics for haptic rendering of basic geometry qne \cite{shen2022mouthhaptics, Jingu2021lipnotif} combines multiple ultrasound haptics for more complex and precise haptic rendering. While many of these papers studied investigated haptics for computer screens \cite{Limerick2019userengagementformidair, Limerick2020calltointeract}, recent work also integrates these haptics for VR applications \cite{Martinez2018TouchlessVRHaptics, Sand2015HMDwithMidAirTactile, Hwang2017AirPiano}. For example, \cite{balint2018medical} demonstrates ultrasound haptics for VR surgical training and \cite{Kervegant2017TouchHologram, Dzidek2018MidAirHapticsInAR} use ultrasound haptics with mixed reality HMDs. 
The main limitation of current ultrasound devices is the limited interaction volume.
Recent work has extended the range of other types of haptic interfaces by integrating them onto robotic arms \cite{Araujo2016SnakeCharmer, Steed2020DockingHaptics}, swarm robots~\cite{suzuki2020roomshift}, or mobile robots \cite{Gonzalez2020REACH+, Suzuki2021HapticBots}.
However To the best of our knowledge, none of the previous work have investigated the combination of ultrasound haptics with robotic actuation.
Taking inspiration from the recent research in AR/VR and robotics~\cite{suzuki2022augmented}, this paper contributes to this new approach that enables the large flexible and scalable, mid-air haptic rendering for AR/VR.
\section{System and Implementation}

\subsection{Hardware}

We'll use Sony Toio \cite{toio} robots for mobile robotic-based 2D translations. Toio robots have two wheels and can move within a 55cm x 55cm area. As shown in Figure 3, we aim to mount ultrasound transducer arrays on top of the Sony Toio robots with 3D printed castors. Sony Toios use built-in cameras to track their position on a printed patterned mat, enabling us to control the x and y positions of the robots with high accuracy. Since the maximum weight capacity of a Sony Toio is about 200g, we 3D print a base with castors to carry the weight of the transducer. We plan to use multiple robots (2-4) for each transducer to enable rapid 2D translations. We plan to build two separate devices so that each hand can be tracked separately and controlled wirelessly controlled by via a Bluetooth enabled PC and Sony Toios. The ultrasound transducers will be controlled via wired connections from a PC.

\begin{figure}[t!]
\centering
\includegraphics[width=1\linewidth]{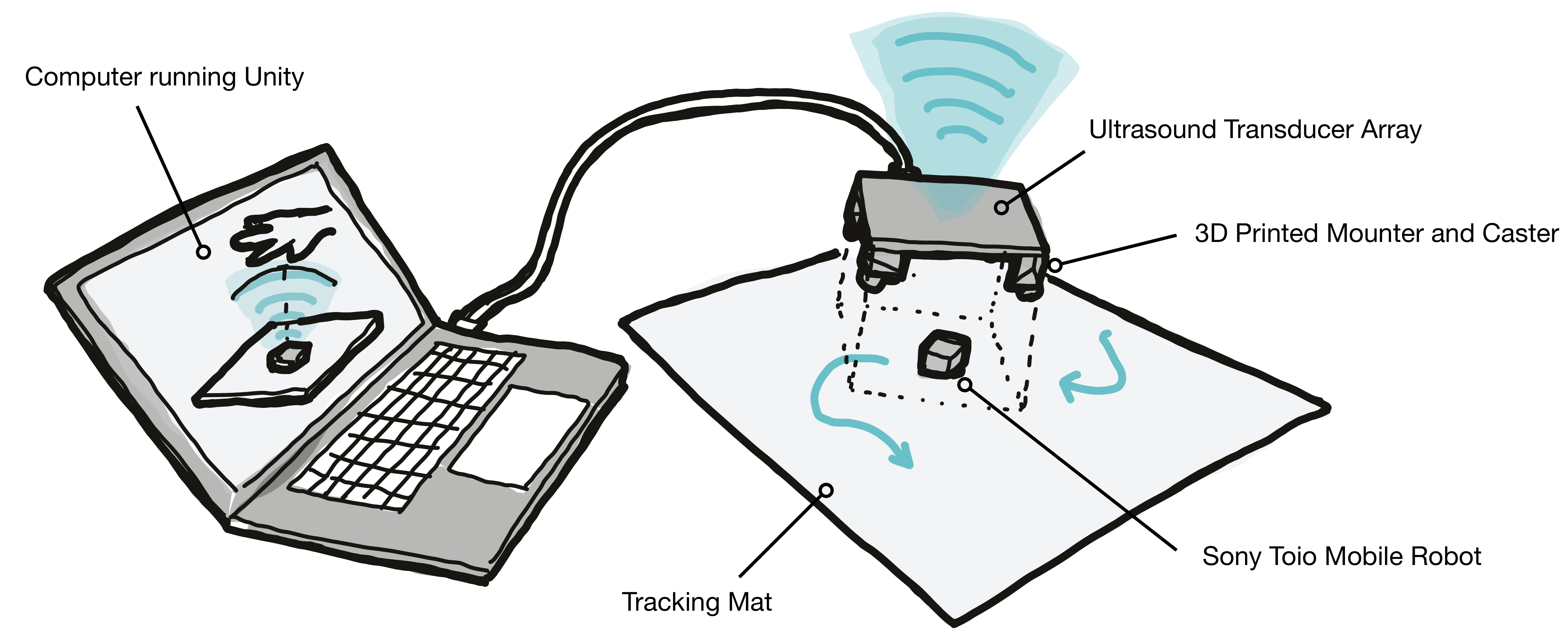}
\caption{System Design to enable 2D translations of ultrasound transducers using Sony Toio tabletop robots}
\label{fig:system-design}
\end{figure}

\subsection{Software}

We create a VR environment using Unity \cite{unity} that synchronizes all of the VR rendering based on user movement, position, hand-tracking, robot control and ultrasound haptics. We communicate with the Toio via a backend Node.js server which provides the position and orientation of each robot to Unity and, based on the virtual objects shape and position, the ultrasound transducer renders the shape of the objects via Unity. The Oculus Quest tracks the users hands and based on that information the system moves the robotic platform carrying the transducers accordingly. For public demonstration purposes, we may investigate a Leap Motion Controller \cite{leapmotion} attached to the robotic platform to be visible and interactable for audiences.

\section{Application}

We implement a medical training application in which the user can touch the human body for massage therapy or surgical training. We also demonstrate an application for a workspace environment, such as keyboard interactions and user interface interactions. Figure 3 displays both medical training and workspace applications. For entertainment purposes, we explore applications such as playing a piano and Whack-a-Mole games.

\begin{figure}[h!]
\centering
\includegraphics[width=0.49\linewidth]{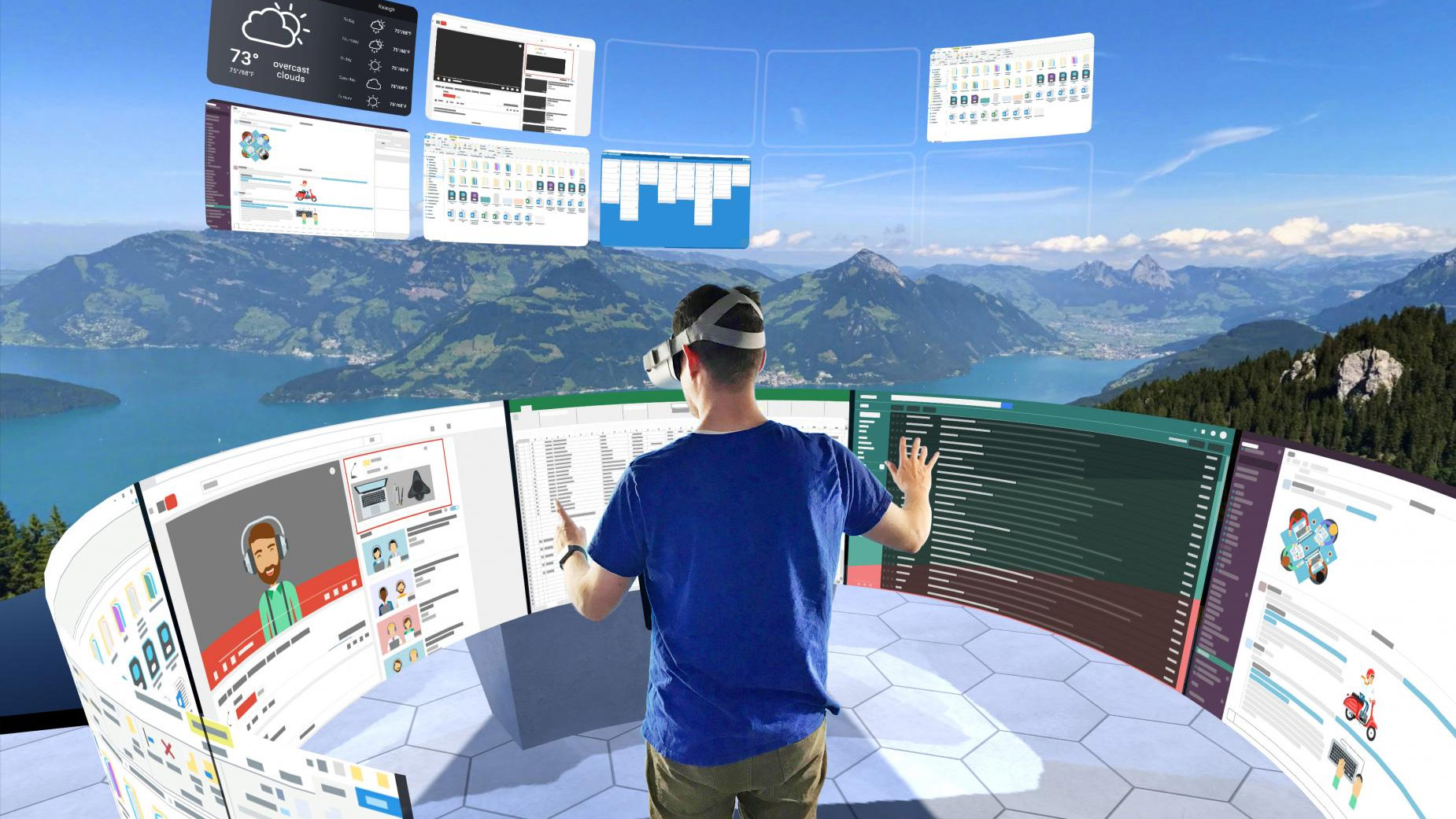}
\includegraphics[width=0.49\linewidth]{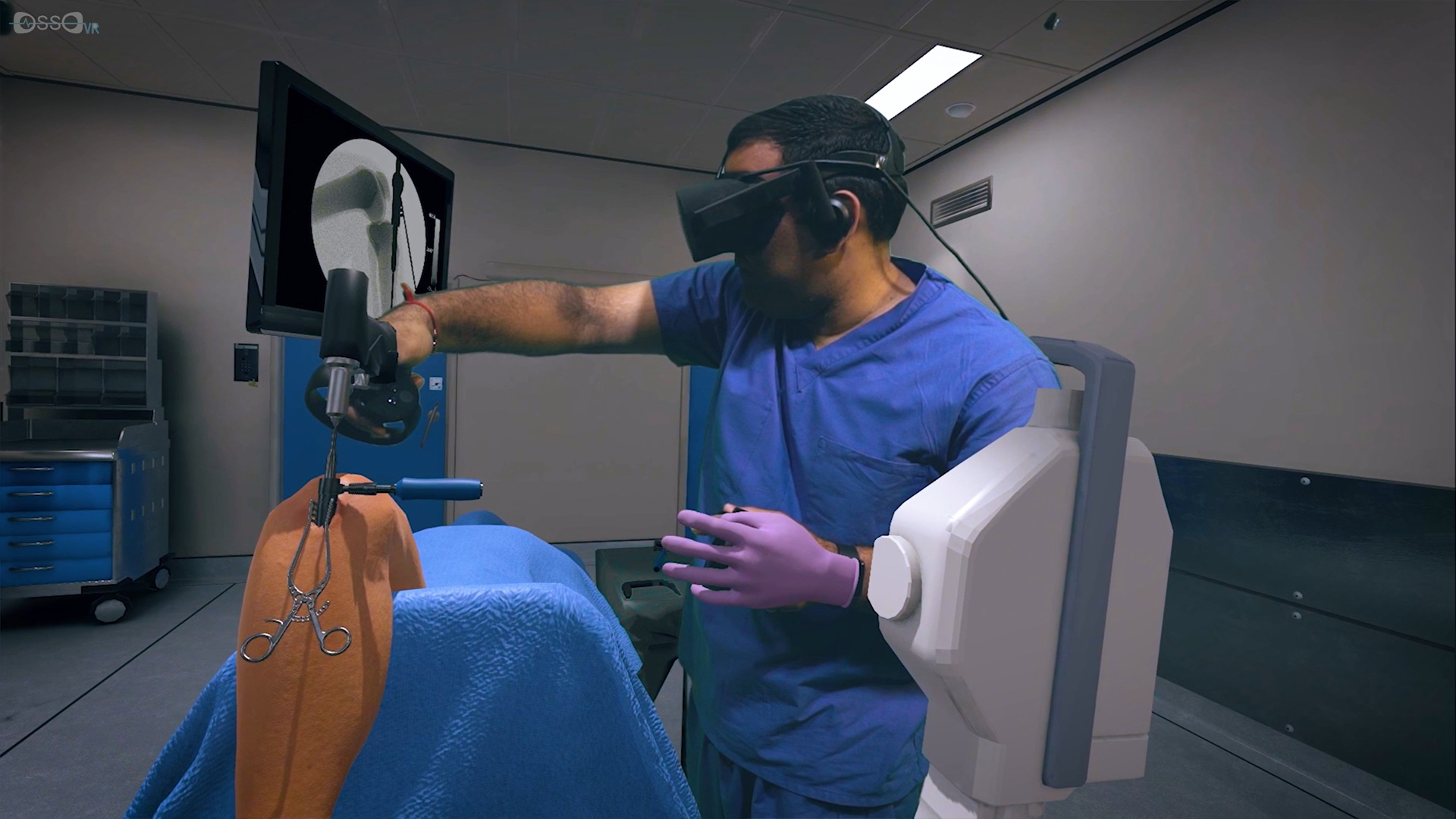}
\caption{Application scenarios. Left: Workspace Environment. Right: Surgical Training.}
\label{fig:applications}
\end{figure}

\ifdouble
  \balance
\fi
\bibliographystyle{ACM-Reference-Format}
\bibliography{references}

\end{document}